\documentclass[12pt]{iopart}

\usepackage{iopams}
\usepackage{epstopdf}
\usepackage{graphicx}
\usepackage{cite}
\usepackage{gensymb}

\begin{document}
\def\be{\begin{equation}}
\def\ee{\end{equation}}
\def\bea{\begin{eqnarray}}
\def\eea{\end{eqnarray}}
\def\rp{r_{+}}
\def\rmm{r_{-}}

\title{Gonihedric (and Fuki-Nuke) Order}

\date{June 2012}
\author{D. A. Johnston}
\address{Dept. of Mathematics and the Maxwell Institute for Mathematical
Sciences, Heriot-Watt University,
Riccarton, Edinburgh, EH14 4AS, Scotland}


\begin{abstract}
A $3D$ Ising model with a purely plaquette, $4$-spin interaction displays 
a planar flip symmetry  intermediate between a global and a gauge symmetry and as a consequence has a highly degenerate low temperature phase and no standard magnetic order parameter. This plaquette Hamiltonian is a particular case of a family of $3D$ Gonihedric Ising models defined by Savvidy and Wegner.  
An anisotropic variant of the the purely plaquette Gonihedric model,
originally discussed as the ``Fuki-Nuke'' model by Suzuki, is non-trivially equivalent to a stack of $2D$ Ising models,
each of which can magnetize independently at the phase transition point.

Consideration of this anisotropic model suggests that a suitable order parameter in the isotropic case may also be constructed using a  form of planar magnetization, in which nearest neighbour correlators $\langle \sigma_i \sigma_j \rangle$ summed over planes replace a sum over spin values $\langle \sigma_i \rangle$. We conduct Monte-Carlo simulations to investigate this and related candidate order parameters in a toy model of Gonihedric ground states, the Fuki-Nuke model and the isotropic plaquette Gonihedric model itself.

\end{abstract} 

\maketitle


\section{Introduction}

The Gonihedric Ising model was defined by Savvidy and Wegner \cite{1} as a cubic lattice
transcription of an earlier Gonihedric random surface model developed by Savvidy
\cite{2}  which was intended to be a discretization of string theory. In the case of the Gonihedric Ising model, the spin cluster boundaries were used to model
a gas of surfaces and the spin couplings were chosen so that edges and intersections of plaquettes rather than their surface areas were weighted, which was also the defining feature of the Gonihedric random surface model. 

Such a weighting can be arranged using a one-parameter family of generalized $3D$ Ising models with fine-tuned nearest neighbour
$\langle ij \rangle$, next to nearest neighbour $\langle \langle ij \rangle \rangle$ and plaquette
interactions $[ijkl]$.  The Hamiltonians in this family are given by
\begin{equation}
\label{e1}
H_{gonihedric} = - 4 \kappa \sum_{\langle ij\rangle }\sigma_{i} \sigma_{j}  +
\kappa \sum_{\langle \langle ij\rangle \rangle }\sigma_{i} \sigma_{j} 
- ( 1-\kappa )\sum_{[ijkl]}\sigma_{i} \sigma_{j}\sigma_{k} \sigma_{l} \; ,
\end{equation}
where the parameter $\kappa$ gives the relative weight of two plaquettes meeting at right angles and four-plaquette intersections in the spin cluster boundaries. 
One member of this family, the purely plaquette action with $\kappa=0$
\begin{equation}
\label{e2k}
H_{\kappa=0} =  -  \sum_{[ijkl]}\sigma_{i} \sigma_{j}\sigma_{k} \sigma_{l}
\end{equation}
has merited particular attention due to its interesting properties, both static and dynamical \cite{3a,3}. It has a
strong first order transition surrounded by a region of metastability and shows evidence of glass-like behaviour at lower temperatures in spite of having no quenched disorder.

The symmetry properties of both the ground states and low temperature phase of $H_{\kappa=0}$ are  unusual. The standard Ising model with only nearest neighbour interactions
\begin{equation}
\label{e0I}
H_{Ising} =  -  \sum_{\langle ij \rangle}\sigma_{i} \sigma_{j}
\end{equation}
displays a twofold symmetry in its ground state, whereas a $3D$ 
$\mathbb{Z}_2$ gauge theory (where the spins live on plaquette edges rather than vertices as in $H_{\kappa=0}$)
\begin{equation}
\label{e0G}
H_{Gauge} =  -  \sum_{[ijkl]} U_{ij} U_{jk} U_{kl} U_{li}
\end{equation}
has the local gauge symmetry $U_{ij} \rightarrow s_i U_{ij} s_j $, with $s_i = \pm 1$. The symmetries of plaquette Gonihedric model
of equ.\,(\ref{e2k}) lie somewhere between these since it is possible to flip arbitrary, possibly intersecting, planes of spins
in the ground state at zero energy cost. Both Monte-Carlo simulations and low temperature expansions \cite{4} confirm that this remains a symmetry of the low temperature phase, which is thus highly degenerate.  A consequence of this symmetry is that the standard magnetization is zero in the low temperature phase. The nature of the magnetic order has thus remained unclear.

In this paper we  discuss the recent suggestion by Hashizume and Suzuki \cite{5} that
a suitable order parameter for $H_{\kappa=0}$ might be constructed using nearest-neighbour spin-spin correlations $\langle \sigma_i \sigma_j \rangle$, which was prompted
by Suzuki's earlier study of an {\it anisotropic} Gonihedric model \cite{6}. We  conduct measurements of this candidate order parameter on 
a toy ensemble of ground state configurations constructed by flipping planes of spins starting from a 
ferromagnetic starting configuration to mimic the ground state structure of the Gonihedric model, before moving on to  simulations of both the anisotropic and isotropic Gonihedric models themselves.

\section{The Anisotropic Gonihedric Model (a.k.a. the Fuki-Nuke Model)}

The study of a strongly anisotropic variant of the plaquette Gonihedric
Hamiltonian by Suzuki \cite{6} pre-dates the work of Savvidy and Wegner by some twenty years.
For reasons that will become apparent, Suzuki dubbed this the ``Fuki-Nuke'' or ``no roof'' model and showed that it was (non-trivially) equivalent to a stack of $2D$ Ising models. This was later rediscovered by Jonsson and Savvidy \cite{7} using a transfer matrix approach and discussed by Castelnovo
{\it et.al.} \cite{8} using methods more akin to Suzuki's original approach.

In Suzuki's  work the principal concern was to obtain solvable $3D$ spin models, and the Fuki-Nuke
model was constructed as a specific example. The starting point was an anisotropically coupled variant of
$H_{\kappa=0}$
\bea
H_{aniso} = - J_x  \sum_{[ijkl]}\sigma_{i} \sigma_{j}\sigma_{k} \sigma_{l}  - J_y \sum_{[ijkl]}\sigma_{i} \sigma_{j}\sigma_{k} \sigma_{l} 
 - J_z \sum_{[ijkl]}\sigma_{i} \sigma_{j}\sigma_{k} \sigma_{l}
\eea
where $J_x, J_y, J_z$ were the couplings for plaquettes perpendicular to the $x,y,z$ axes respectively. 
Setting $J_x=J_y=1$ and $J_z=0$ gave a model where the horizontal ``roof'' plaquettes had zero coupling, hence the sobriquet of ``Fuki-Nuke''. 

To show that this is equivalent to a stack of $2D$ Ising models requires a little further work, which can be done succinctly by using the non-linear $\sigma$-$\tau$ transformation, following Suzuki \cite{6} and  Castelnovo {\it et.al.} \cite{8}. We first define bond spin variables $\tau_{j} = \sigma_i \sigma_j$ that can be thought of as living at the centre of (only) the vertical lattice bonds.  
Since we have condensed the site labels on both sets of spins, we  write $\sigma_k$ and $\tau_k$ to represent any of the spins in the $kth$ horizontal $2D$ layer. With this notation the direct transformation for a spin in the $kth$  layer  is
\be 
\sigma_k = \tau_1 \tau_2 \tau_3 \cdots \tau_k
\ee
and the inverse transformation is given by 
\be
\tau_1 = \sigma_1, \;  \tau_2 = \sigma_1 \sigma_2 \; , \; \ldots \; , \; \tau_N = \sigma_{N-1} \sigma_N
\ee
In order to obtain a one-to-one correspondence between
$\sigma$ and $\tau$ spin configurations we need to specify the value of the $\sigma_1 \, / \, \tau_1$ spins on a given (first in this case) horizontal plane,
but in all the other planes a product of two $\sigma$ spins at the end of each
vertical bond gives the $\tau$ spin.  
The resulting Hamiltonian is then
\be
H = - \sum_i \left( \tau_{i}  \tau_{i + \hat e_x} + \tau_{i} \tau_{i + \hat e_y} \right)
\ee
where $\hat e_x, \hat e_y$ are unit vectors in the horizontal directions.
We have not written the contribution of the fixed plane, which will vanish in the thermodynamic limit. The $\tau_{i}$ spins are coupled to their horizontal nearest neighbours in the ${\hat e_x}$ and ${\hat e_y}$ directions
and there are  no vertical inter-plane couplings.
We have thus shown that the Fuki-Nuke model is equivalent to a stack of
uncoupled $2D$ Ising models arranged horizontally one above each other.

The magnetic order parameter for the $i$th $2D$ Ising layer is
given by the standard expression
\be
\label{Mone}
M_{2D, \, i} =  \left< \frac{1}{N^2} \sum_{single \; plane} \tau_{i} \right>
\ee
which may be rewritten in terms of the original $\sigma_i$ spins as
\be 
M_{2D, \, i} = \left<  \frac{1}{N^2} \sum_{single \; plane}   \sigma_i \sigma_{i +  \hat e_z}   \right> \; .
\ee
where $\hat e_z$ is  the unit vector in the vertical direction. The nearest neighbour $ \sigma_i \sigma_{i +  \hat e_z}$ correlator is measured on all the vertical bonds bisected by the horizontal plane of the corresponding $\tau_i$ spins. 
The unusual appearance of a two-spin correlator as an order parameter is simply the result
of expressing the magnetization in terms of the original $\sigma_i$ variables rather than the bond spins $\tau_{i}$. Remembering that
$\tau_i =  \sigma_i \sigma_{i +  \hat e_z}$ and using the standard 
results for the $2D$ Ising model magnetization   we have \cite{6}
\be
\label{M2D} 
M_{2D, \, i} = \left<  \frac{1}{N^2} \sum_{single \; plane} \sigma_i \, \sigma_{i +  \hat e_z}  \right>  = \pm \left( 1 - \sinh^{-4}( \beta)\right)^{1/8} 
\ee
which will behave as $\pm \left| \beta - \beta_c \right|^{1 \over 8}$
near the critical point $\beta_c = \ln ( 1 + \sqrt{2})$.
More generally, since
$\sigma_{k} = \tau_1 \, \tau_2 \, \ldots \, \tau_{k}$ and the different $\tau_i$ layers are decoupled,
\be 
M_{2D, \, i, \, n} = \left<\frac{1}{N^2} \sum_{single \; plane} \sigma_i \, \sigma_{i +  n \hat e_z} \right> =   ( M_{2D, \, i} )^n \; .
\ee
The in-plane horizontal correlations between ``dimers'' of $\sigma_i$ spins,
where the $i,j$ spins now lie on the same plane,
\be 
\left< ( \sigma_i \, \sigma_{i +  \hat e_z} ) (  \sigma_{j} \, \sigma_{j+  \hat e_z}) \right> = \left< \tau_{i} \, \tau_{j} \right> 
\ee
are also, of course, Ising-like.

Since the various  $2D$ Ising layers magnetize independently  we could construct a quasi-$3D$ order parameter to be
of the form
\be
\label{Mabs}
M_{abs}  = \frac{1}{N} \sum_{xy \, planes} \Big\langle  \Big| \frac{1}{N^2} \sum_{single \; plane}   \sigma_i \sigma_{i +  \hat e_z}  \Big| \Big\rangle 
\ee
or 
\be
\label{Msq}
M_{sq} =  \frac{1}{N} \sum_{xy \, planes} \Big\langle  \left( \frac{1}{N^2}  \sum_{single \; plane}   \sigma_i \sigma_{i +  \hat e_z}   \right)^2 \Big\rangle \; 
\ee
to avoid inter-plane cancellations.
We have explicitly retained the various normalizing factors in equs.(\ref{Mabs}, \ref{Msq}) for a cubic  lattice with $N^3$ sites. 
We could also define similar quantities for non-nearest neighbour correlators
\be 
\label{MInabsxyz}
M_{abs,\, n}^{} =  \frac{1}{N} \sum_{xy \, planes} \Big\langle  \Big| \frac{1}{N^2} \sum_{single \; plane}   \sigma_i \sigma_{i + n \hat e_{z} } \Big| \Big\rangle
\ee
or
\be 
\label{MInsqxyz}
M^{}_{sq, \, n} =  \frac{1}{N} \sum_{xy  \, planes} \Big\langle  \left( \frac{1}{N^2}  \sum_{single \; plane}  
 \sigma_i \sigma_{i +  n \hat e_{z}}   \right)^2 \Big\rangle  
\ee
with $n>1$. Since each plane magnetizes at the same point the resulting expressions are given by
\begin{eqnarray}
\label{Msinh}
M_{abs} &=& \left( 1 - \sinh^{-4}( \beta)\right)^{1/8}  \nonumber \\
M_{sq} &=&  \left( 1 - \sinh^{-4}( \beta)\right)^{1/4}  \nonumber \\
M_{abs, \, n} &=& \left( 1 - \sinh^{-4}( \beta)\right)^{n/8}  \nonumber \\
M_{sq, \, n} &=&  \left( 1 - \sinh^{-4}( \beta)\right)^{n/4}  \nonumber \; .
\end{eqnarray}
As the various other variants of the magnetization are given in terms of $M_{abs} = | M_{2D, \, i} |$ this is probably the most
suitable  choice of ``three-dimensional'' order parameter for the Fuki-Nuke model.  
In addition, it emphasizes that  the critical behaviour of the Fuki-Nuke model is really that of the $2D$ Ising model in each layer. 

In the anisotropic model the orientation of the planes in the summation is fixed by the choice of zero coupling, which we have taken to be horizontal planes in our case.  The suggestion by Hashizume and Suzuki  in \cite{5}  is that a similar order parameter measuring coplanar, nearest neighbour spin correlations could still serve for the {\it isotropic} Gonihedric model.
They considered a mean-field approach to the model and correlation function inequalities in order to corroborate this and found supporting evidence from both, as well as extending the discussion to disordered couplings. 
Here, we 
investigate the order parameter in the isotropic $3D$ Gonihedric model  directly using Monte-Carlo simulations, restricting discussion to the purely ferromagnetic case for simplicity.

In equs.\,(\ref{Mabs},\ref{Msq}) we have introduced either a modulus or square for the spin correlators in order to avoid inter-plane cancellations, whereas Suzuki's approach is tantamount to considering the direct  equivalent of equ.\,(\ref{Mone})
\be 
\label{MSuz}
M =  \frac{1}{N} \sum_{planes} \Big\langle   \frac{1}{N^2} \sum_{single \; plane}   \sigma_i \sigma_{i +  \hat e_z}   \Big\rangle 
\ee
in the presence of an external field (later taken to zero) which picks out one of the $2D$ magnetizations on each plane. 
The possibility of using an order parameter akin to the $M_{sq}$ in equ.\,(\ref{Msq}) has been  mentioned previously by Lipowski \cite{3a}, who noted in a simulation of $H_{\kappa=0}$ that it  appeared to possess the correct behaviour. He also remarked on the possibility of using non-nearest neighbour correlators as the order parameter(s), which we shall discuss further below.
In the remainder of the paper we compare measurements of the candidate order parameters on toy ground state configurations, the Fuki-Nuke model and the isotropic $3D$ Gonihedric model itself.

\section{Monte Carlo investigations: Toy Model, Fuki-Nuke and the Real Thing}

\subsection{Toy Model - Ground States}
Ground states of the  Gonihedric Hamiltonian $H_{\kappa=0}$  may be obtained by flipping some set of planes of spins in a reference ferromagnetic configuration which has all spins up or down. A simple test for any candidate order parameter is then to generate an ensemble of such configurations and measure its expectation value. A typical configuration on a finite cubic lattice
generated in this manner would look something like that in Fig.\,(1).
\begin{figure}[h]
\begin{center}
\includegraphics[height=4cm]{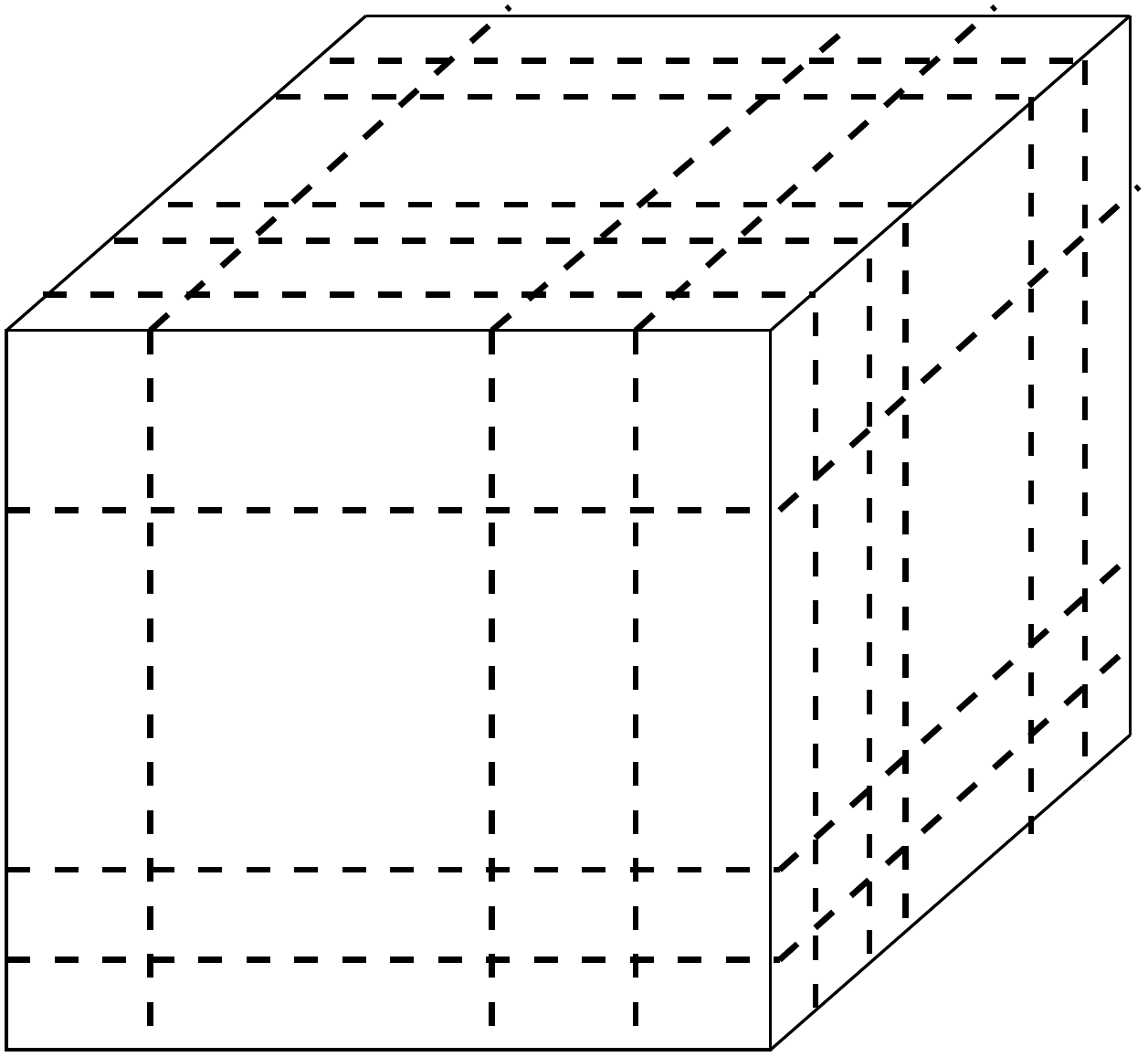}
\label{cube_stripe} 
\caption{A ``ground-state'' configuration generated by flipping arbitrary planes of spins, shown dotted, 
starting in a ferromagnetic state with all the spins positive or negative.}
\end{center}
\end{figure}
We can also measure proposed order parameters on purely random (i.e. infinite temperature) configurations. Any that fail to distinguish between the 
 ground states and random configurations would not, at first sight, appear to be suitable  for use in a  simulation of the full Hamiltonian.

If the suggestion of \cite{5} is correct, we would expect that the order in the
low temperature phase of the  
isotropic $3D$ Gonihedric model could be discerned using an appropriate generalization of either the  $M_{abs}$ and $M_{sq}$ in equs.\,(\ref{Mabs},\ref{Msq}). For   isotropic  configurations such as that in Fig.\,(1) the system should be agnostic to the orientation
of the planes used in calculating the correlators, so any of the three choices of orientation for reference planes in
\bea 
\label{Mabsxyz}
M^{x,y,z}_{abs} &=&  \frac{1}{N} \sum_{yz|xz|xy \; planes} \Big\langle  \Big| \frac{1}{N^2} \sum_{single \; plane}   \sigma_i \sigma_{i +  \hat e_{x,y,z}} \Big| \Big\rangle 
\eea
or
\bea
\label{Msqxyz}
M^{x,y,z}_{sq} &=&  \frac{1}{N} \sum_{yz|xz|xy  \, planes} \Big\langle  \left( \frac{1}{N^2}  \sum_{single \; plane}  
 \sigma_i \sigma_{i +  \hat e_{x,y,z}}   \right)^2 \Big\rangle  
\eea
where $\hat e_x, \hat e_y, \hat e_z$ are the unit vectors in the $x,y,z$ directions respectively,
should be able to serve as order parameters. In all cases the correlators are nearest neighbour correlators for the $\sigma_i$ spins on bonds perpendicularly
bisected by the reference planes, as in the Fuki-Nuke model.
Reassuringly, we find that the various  $M^{x,y,z}_{abs}$ from equ.\,(\ref{Mabsxyz}) and the $M^{x,y,z}_{sq}$ from equ.\,(\ref{Msqxyz}) all give $+1$
when measured on an ensemble of flipped states such as that in Fig.\,(1) so they do, indeed, pick up the planar order
that is characteristic of the Gonihedric models. In addition, the $M^{x,y,z}_{abs}$  and $M^{x,y,z}_{sq}$ all give zero when measured on purely random configurations, so they are capable of distinguishing this order from the complete disorder at infinite temperature.

The choice of nearest neighbour correlators in defining the order parameter does not  appear to be obligatory, though given the line of reasoning leading to it from the Fuki-Nuke model it is probably the most natural. If we measure quantities such as
\be 
\label{Mnabsxyz}
M_{abs,\, n}^{x,y,z} =  \frac{1}{N} \sum_{yz|xz|xy \, planes} \Big\langle  \Big| \frac{1}{N^2} \sum_{single \; plane}   \sigma_i \sigma_{i + n \hat e_{x,y,z}}  \Big| \Big\rangle
\ee
or
\be 
\label{Mnsqxyz}
M^{x,y,z}_{sq, \, n} =  \frac{1}{N} \sum_{yz|xz|xy  \, planes} \Big\langle  \left( \frac{1}{N^2}  \sum_{single \; plane}  
 \sigma_i \sigma_{i +  n \hat e_{x,y,z}}   \right)^2 \Big\rangle  
\ee
with $n>1$ on the ensemble of flipped ground state configurations these also give $+1$ (and zero on random configurations). That this should be so is clear from starting with configurations in which planes of one orientation only, say horizontal, are flipped. For these, plane-to-plane correlations at any separation must be $\pm 1$. Taking a modulus or square for each planar sum then makes all of these $+1$. Flipping planes of spins in the two remaining directions perpendicular to this will change the signs of lines of spins
on both the horizontal planes contributing to the correlator sums and therefore not change the values of the contributing correlators.

When the standard magnetization is zero
\be 
M = \frac{1}{N^3} \sum_i \langle \sigma_i \rangle =0  
\ee
the magnetic susceptibility $\chi$ is given by
\be 
\label{chi}
\chi = \frac{1}{N^3} \sum_i \sum_j \langle \sigma_i \sigma_j \rangle = \frac{1}{N^3} \sum_i \sum_n \langle \sigma_i \sigma_{i+n} \rangle
\ee 
which is  related to the class of order parameters discussed here, since the sums may be arranged in a similar manner to those in $M_{sq}$ and $M_{abs}$. However, the expression for $\chi$ includes a summation over distances greater than one and also retains the signs of the various different contributions in the sum.
Consequently, the susceptibility does not distinguish between configurations such as those in Fig.\,(1) and random configurations, giving $+1$ in both cases, but as we shall see below  it  {\it does} appear to display the characteristics of an order parameter
in a simulation of the full isotropic Gonihedric model.

\subsection{Fuki-Nuke Model}
Since the anisotropic Gonihedric Hamiltonian, or Fuki-Nuke model, originally inspired the definition of the candidate order parameters it too can  serve as a test case for simulations. We know  that $2D$ Ising criticality is expected in each horizontal layer, so the transition point will be seen in the continuum limit at the $2D$ Ising value of $\beta_c = \ln ( 1 + \sqrt{2} ) \simeq 0.88$. 
In the Fuki-Nuke model the orientation of the the $2D$ planes used to measure the nearest neighbour spin correlations is fixed by the choice of the zero couplings. In our simulations we take $J_z=0$, which gives horizontal planes in the sums
and means that $M^z_{abs}$ should be the correct choice for the order parameter.

We simulate the Fuki-Nuke model on a $20^3$ lattice with periodic boundary conditions using $10^7$ Metropolis Monte-Carlo measurement sweeps with a hot start and an equilibration time of $10^5$ sweeps as a test case.
In Fig.\,(2) we can see that both $M_{abs}$ and $M_{sq}$ give a clear signal for the pseudo-critical point, which can be observed close to the continuum value of  
$\beta \sim 0.88$ already on  the $20^3$ lattice via, for example, the peak in the specific heat measurements.  As a consequence of cancellations between differently signed $2D$ layers of Ising spins in the magnetized phase  
the standard magnetization $M$ remains zero for all $\beta$ in the Fuki-Nuke model
and does not give an obvious  signal for the transition. 
\begin{figure}[h]
\begin{center}
\includegraphics[height=6cm]{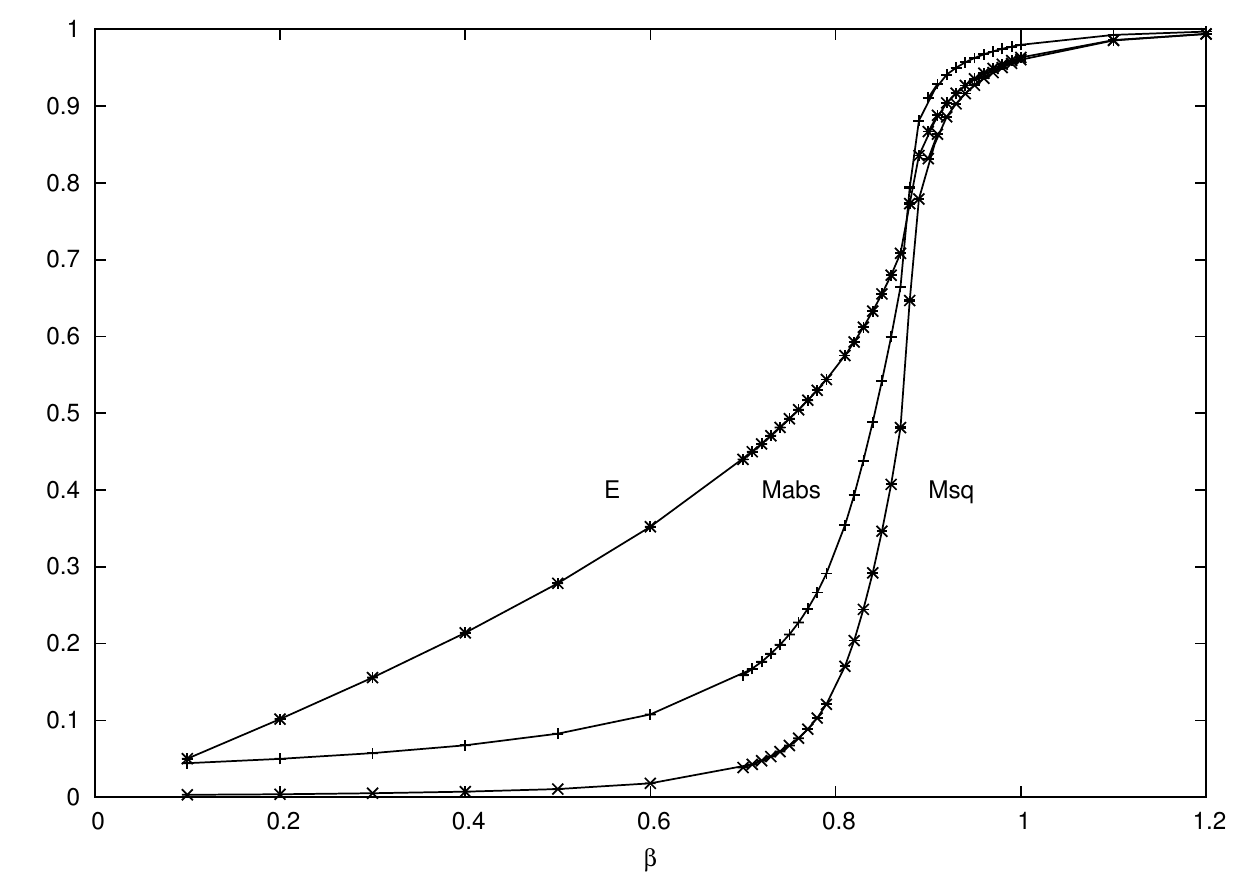}
\label{aniso20} 
\caption{The energy and order parameters $M_{abs}$ and $M_{sq}$ on a $20^3$  lattice for the Fuki-Nuke model.}
\end{center}
\end{figure}

\subsection{(Isotropic) Gonihedric Model}

The preceding measurements reported for the toy ground state model and the Fuki-Nuke model suggest that
any of $M^{x,y,z}_{abs}$ from equ.\,(\ref{Mabsxyz}), the  $M^{x,y,z}_{sq}$ from equ.\,(\ref{Msqxyz}) or even the $M^{x,y,z}_{abs, \, n}$ and
$M^{x,y,z}_{sq, \, n}$ of equs.\,(\ref{Mnabsxyz},\ref{Mnsqxyz}) might be viable candidates for an order parameter in the isotropic $3D$ Gonihedric model itself.
To clarify this, we  carried out simulations 
of the  isotropic Gonihedric model on $10^3$ and $15^3$ lattices using
$10^7$ Metropolis Monte-Carlo measurement sweeps, following a hot start and an equilibration time of $10^5$ sweeps. The various $M^{x,y,z}_{abs}$ and $M^{x,y,z}_{sq}$  along with
$M^{x,y,z}_{abs, \, n}$ and $M^{x,y,z}_{sq, \, n}$ for $n>1$ were measured on both lattices.

In Fig.\,(3) we plot the measurements of $M_{sq}^{x,y,z}$ and $M_{abs}^{x,y,z}$
on the $10^3$ lattice and in Fig.\,(4) on the $15^3$ lattice. The measurements of the $M$'s for the different $x,y,z$ orientations of the reference planes are indistinguishable from each other and the observed jump in the various $M$'s coincides with the pseudo-critical values of $\beta$ determined from the jump in the energy observed
in simulations ($\sim 0.54$ on the $10^3$ lattice and $\sim 0.57$ on the $15^3$ lattice with ``hot'' starts).  These results confirm that the orientation of the reference planes is, indeed, irrelevant in measuring these magnetizations and that all three choices are equivalent. They also show clearly that the 
correlators distinguish between the disordered and ordered phases of the isotropic model as suggested by the arguments in \cite{5}.
\begin{figure}[h]
\begin{center}
\includegraphics[height=6cm]{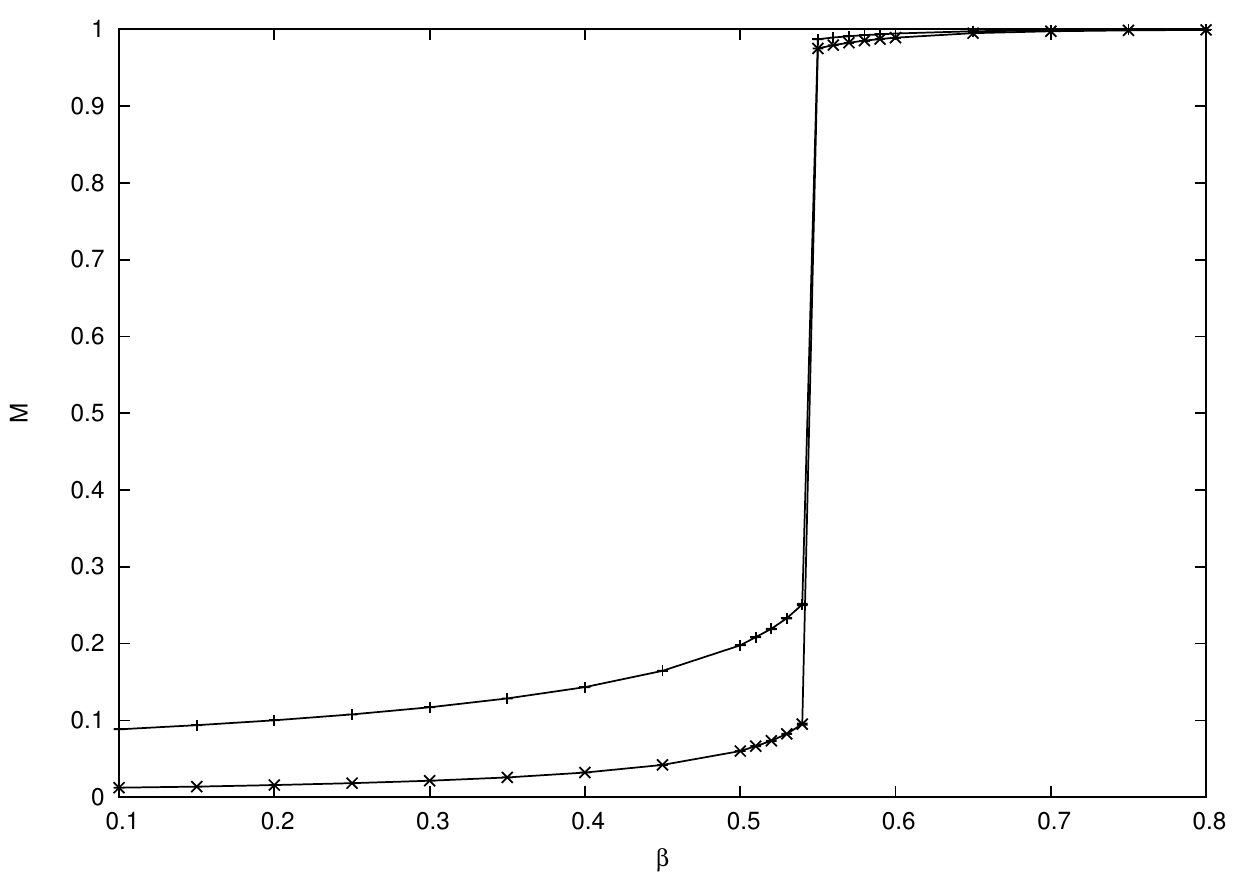}
\label{M_10} 
\caption{$M_{abs}^{x,y,z}$ (upper) and $M_{sq}^{x,y,z}$ (lower) on a $10^3$ lattice for the isotropic Gonihedric model. Lines are drawn to guide the eye.}
\end{center}
\end{figure}
\begin{figure}[h]
\begin{center}
\includegraphics[height=6cm]{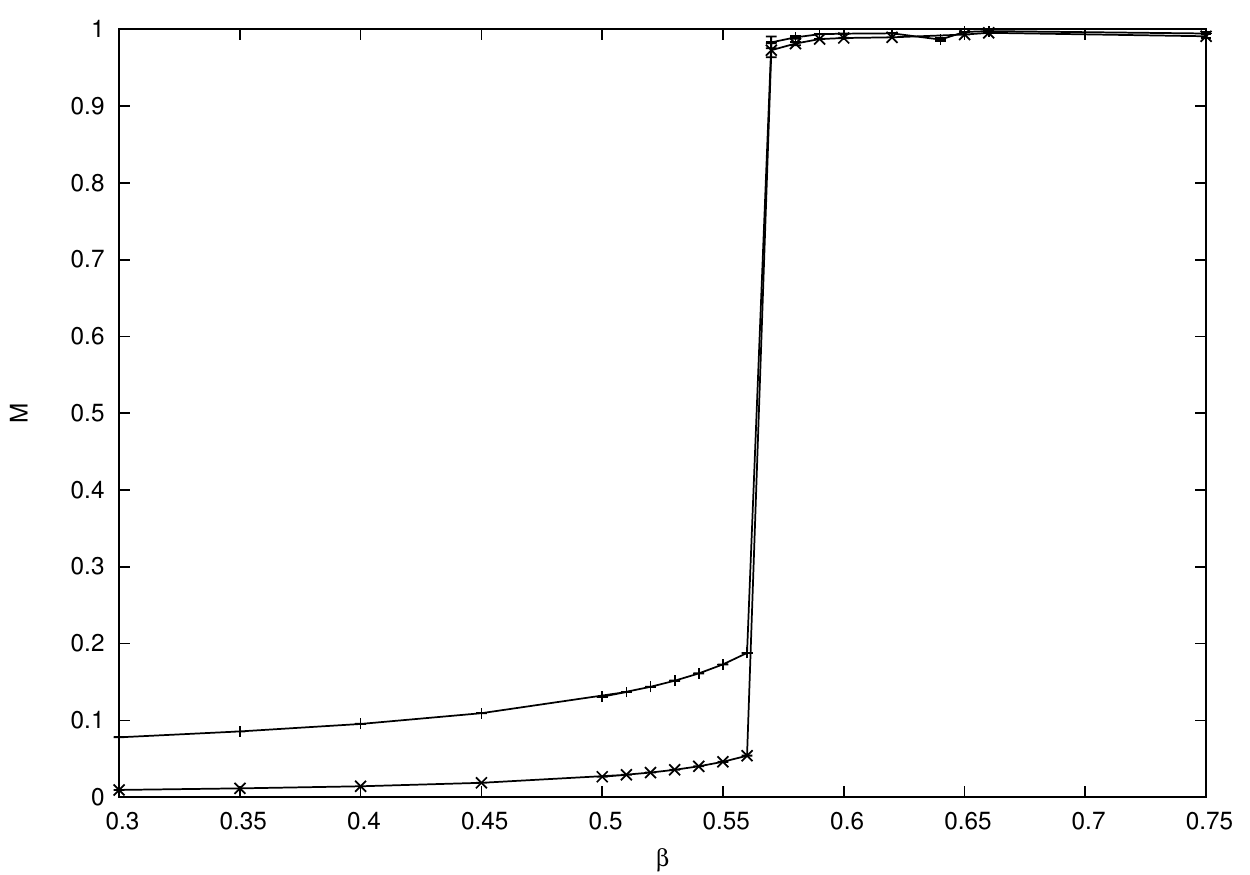}
\label{Msq_20} 
\caption{$M_{abs}^{x,y,z}$ (upper) and $M_{sq}^{x,y,z}$ (lower) on a $15^3$ lattice for the isotropic Gonihedric model. Again, lines are drawn to guide the eye.}
\end{center}
\end{figure}

To guard against the possibility of discarding too much information by considering a simple global order parameter we can measure the distribution  $P(M_{2D, \,  i})$ of the coplanar $2D$ spin correlations. In Figs.\,(5,6)
we have combined the three orientations of reference planes to plot a $P(M_{2D})$  in the high temperature and low temperature phases respectively. The three individual orientations contributing to this give identical distributions. In Fig.\,(5) $P(M_{2D})$ is  gaussian (narrowing with system size and increasing temperature) and centred 
at $M_{2D}=0$, whereas in the low temperature phase in Fig.\,(6) at $\beta=0.58$ there are two sharp peaks at $M_{2D} = \pm 1$. We would expect the slight asymmetry between the peaks in the plot to disappear with better statistics and increasing lattice size.
The observed behaviour strongly supports the hypothesis of Fuki-Nuke type order, where coplanar correlators ``magnetize'' independently of the other planes, in the isotropic model too.
The plots also demonstrate that a $[0,1]$ type order parameter is sufficient for the isotropic Gonihedric model, since there is no non-trivial structure in $P(M_{2D})$ such as that seen in the overlap distribution $P(q)$ of (mean-field) spin glasses.   
\begin{figure}[h]
\begin{center}
\includegraphics[height=5cm]{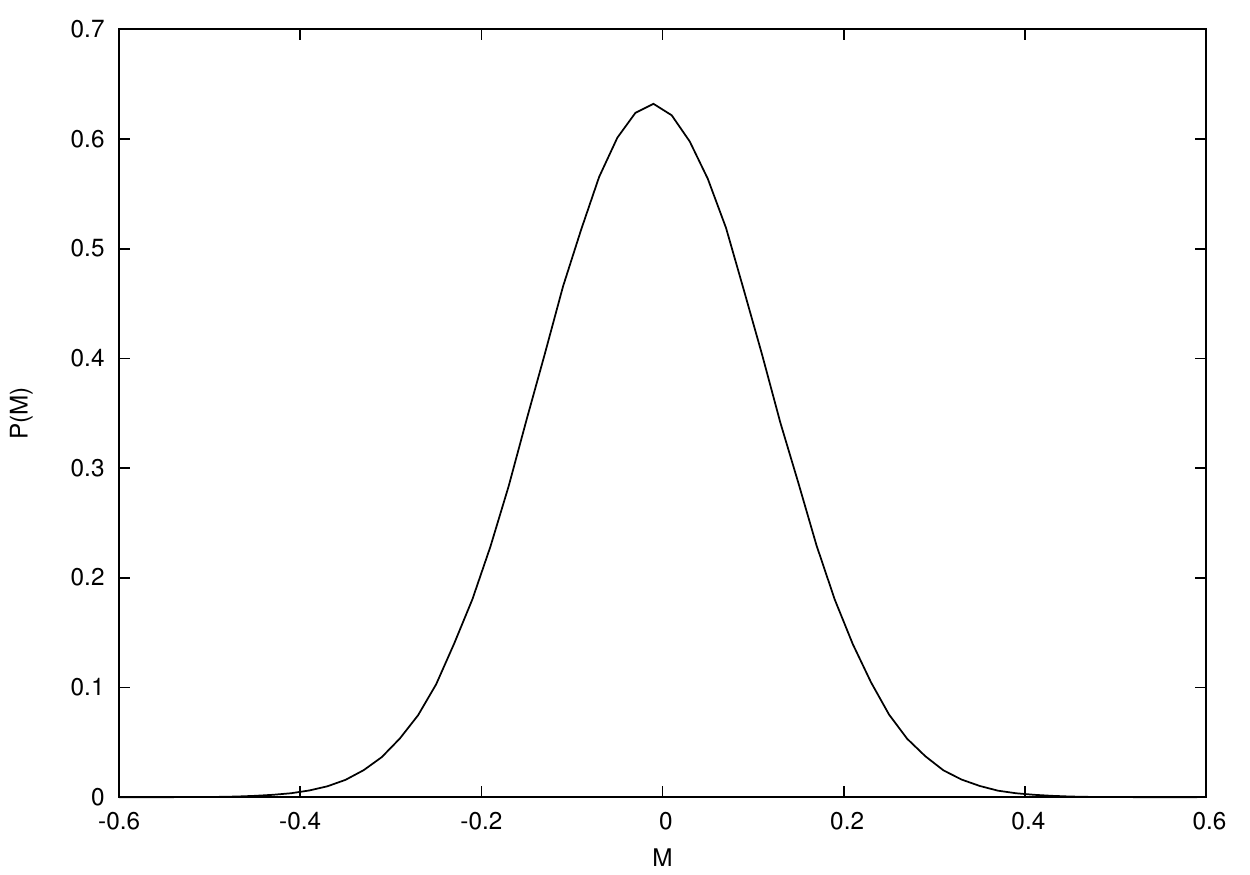}
\label{H_10} 
\caption{$P(M_{2D})$ at $\beta=0.20$ in the high temperature phase on a $10^3$ lattice}
\end{center}
\end{figure}
\begin{figure}[h]
\begin{center}
\includegraphics[height=5cm]{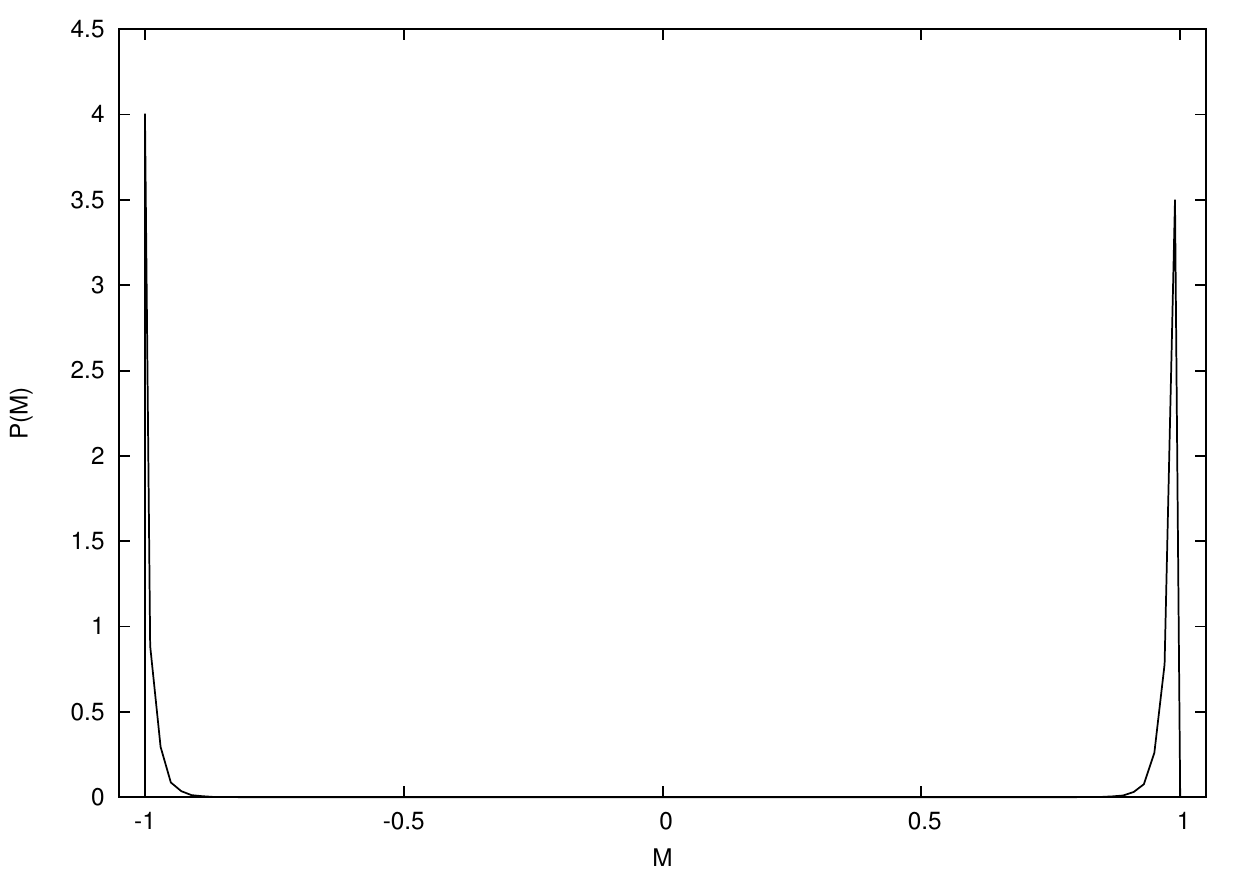}
\label{H_58} 
\caption{$P(M_{2D})$ at $\beta=0.58$ in the low temperature phase on a $10^3$ lattice}
\end{center}
\end{figure}

We also find, as with the toy ground state ensemble, that $M_{abs, \, n}^{x,y,z}$ and
$M_{sq, \, n}^{x,y,z}$ for $n>1$ show very similar behaviour to $M_{abs}^{x,y,z}$ and
$M_{sq}^{x,y,z}$.
In Fig.\,(7)  $M_{abs, \, 6}^z$ on a 
$15^3$ lattice is plotted, along with $M_{abs}^z$ for comparison. $M_{abs, \, 6}^z$ is typical of the various $M_{abs, \, n}$ (and $M_{sq, n}$) for other $n$.
\begin{figure}[h]
\begin{center}
\includegraphics[height=6cm]{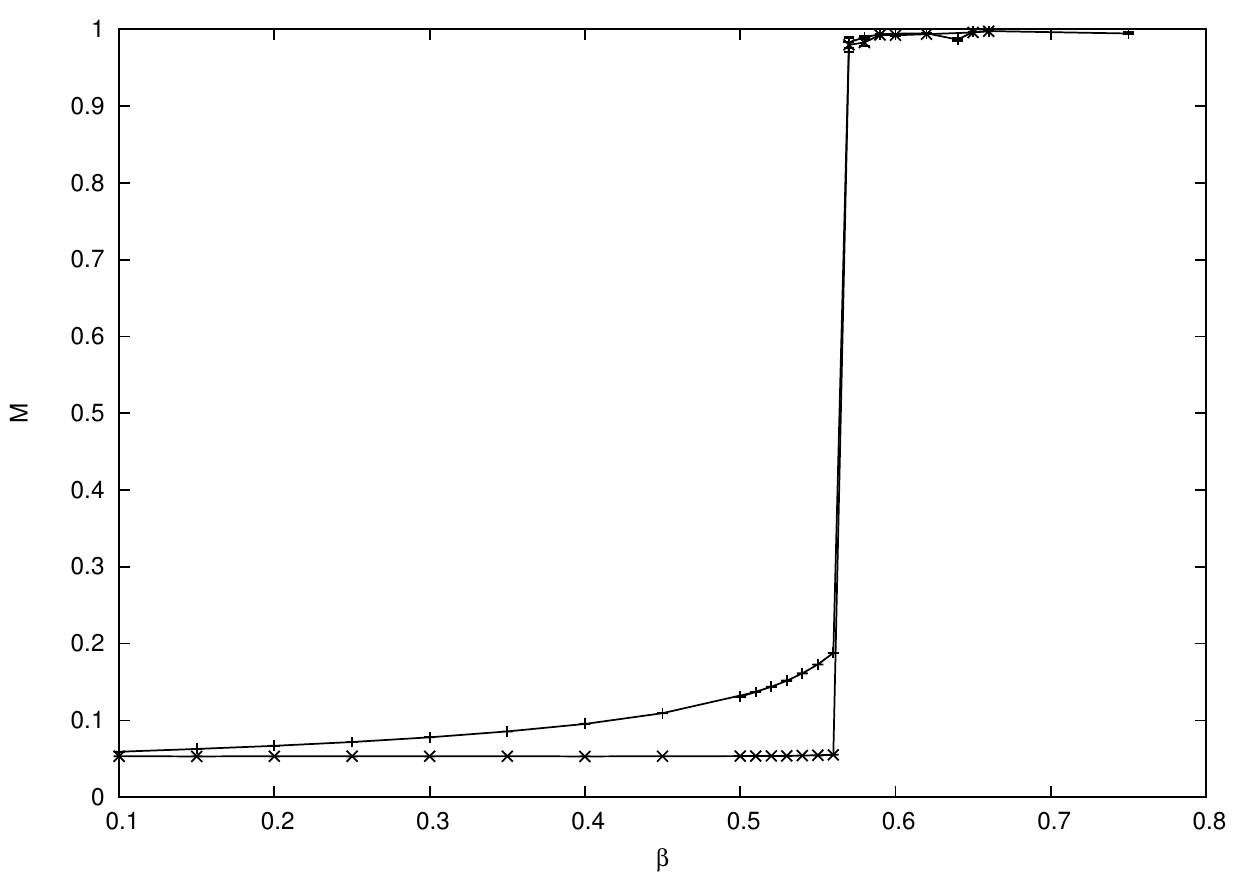}
\label{Mabs_15_06} 
\caption{$M_{abs}^{z}$ (upper) and $M_{abs, 6}^{z}$ (lower) on a $15^3$ lattice for the isotropic Gonihedric model.}
\end{center}
\end{figure}
As might be expected the magnetization constructed from the more widely spaced correlators in $M_{abs, 6}^{z}$ is lower in the disordered phase, but it too jumps sharply to one at the transition point just as does $M_{abs}^{z}$.  Since we employ periodic boundary conditions the measured magnetizations repeat from $M_{abs, 7}$ on the $15^3$ lattice as  the distances in the correlators wrap around the lattice. With the isotropic model there is no longer a mapping to the $2D$ Ising model, unlike the Fuki-Nuke model, so we do not have an exact expression for the planar average of nearest neighbour correlators as in equ.\,(\ref{M2D}), nor do we have any assurance that different (bi-)layers of spins decouple, which allows us to express everything in terms of the nearest neighbour correlators alone in the Fuki-Nuke case. Nonetheless, it would appear that the most fundamental correlations in the isotropic case are also those between nearest neighbour spins. The observed signal for the transition in the higher correlators are a consequence of that in the nearest neighbour correlations, just as with the toy ground states and the Fuki-Nuke model.  

A curious feature of the isotropic Gonihedric model simulations that should be highlighted is that
the standard magnetic susceptibility, as defined in equ.\,(\ref{chi}):
$\chi = \frac{1}{N^3} \sum_i \sum_j \langle \sigma_i \sigma_j \rangle = \frac{1}{N^3} \sum_i \sum_n \langle \sigma_i \sigma_{i+n} \rangle$,
also behaves as an effective order parameter, unlike the toy ground state ensemble. This can be seen in Fig.\,(8) where $\chi$ is  $+1$  at high temperatures and drops sharply to zero 
at the observed pseudo-critical points. 
\begin{figure}[h]
\begin{center}
\includegraphics[height=6cm]{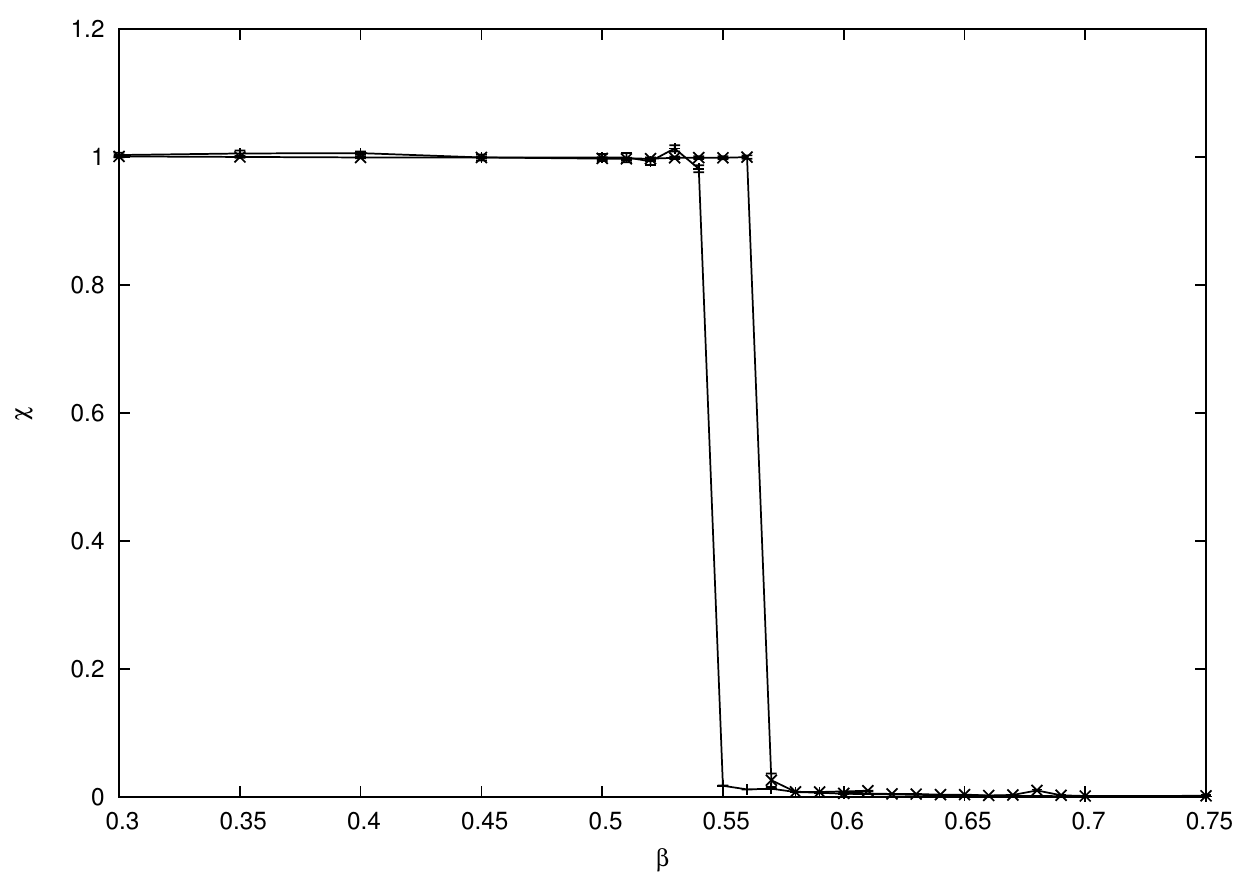}
\label{Susp_10_20} 
\caption{The magnetic susceptibility $\chi$ on both the $10^3$ and  $15^3$ lattices is  one at high temperatures and drops sharply to zero 
at the observed pseudo-critical points.}
\end{center}
\end{figure}
We have already seen in our discussion of the toy model for ground states that $\chi$ does {\it not} distinguish between random, infinite temperature configurations and the 
flipped ground-state configurations, which means that the observed behaviour is not  a direct consequence of $\chi$ detecting
the coplanar order of the low temperature phase as with $M^{x,y,z}_{abs,sq}$. It appears rather to be a dynamical  ergodicity breaking  effect. At high temperatures the flip symmetry of the Hamiltonian is not broken, so $\chi$ must be one, whereas the observed value of zero for $\chi$ throughout the  low temperature (high $\beta$) phase is due to the  system ``freezing'' into one configuration as the Metropolis spin flip acceptance drops rapidly to zero at $\beta_c$. It would be interesting to explore the behaviour of $\chi$ in a multicanonical simulation such as that  
carried out in \cite{Marco}. In this case the system would still be free to explore multiple ground states, so the behaviour of $\chi$ would presumably be more akin to that in the toy ground state ensemble (and less like an order parameter).

It is also worth noting that the strong hysteresis associated with the first order transition means that  a more accurate estimate of the continuum transition
point $\beta_c (\infty)$ is best done with such a multi-canonical simulation, or similar. In \cite{Marco} the value of  $\beta_c (\infty) = 0.549 25(6)$ obtained from the finite size scaling of the specific heat maxima is a little higher than that obtained 
 from taking careful account of scaling corrections arising from  the boundary conditions,  $\beta_c (\infty) = 0.547 57(63)$,
using a Metropolis simulation \cite{Fixed}. The various pseudo-critical temperatures observed on differently sized lattices in a multi-canonical simulation fall between those observed for hot and cold starts in Metropolis simulations such as those performed here on the same lattices, as one might expect.

\section{Conclusions} 
 
As Suzuki remarked many years ago \cite{6} the anisotropic Gonihedric Ising (Fuki-Nuke) model may be
reformulated in terms of new spin variables as a stack of decoupled $2D$ Ising models which magnetize independently. 
When translated back into the original spin variables the magnetization of each $2D$ Ising model is expressed in terms of a 
two spin nearest neighbour correlator $\langle \sigma_i \sigma_j \rangle$.
To take account of the independent ordering of planes a  quasi-$3D$ order parameter may then be constructed by taking  the modulus, $M_{abs}$ 
of nearest neighbour correlators summed over  each plane.

We investigated numerically 
the suitability of using this and related quantities as an order parameter for the  isotropic $3D$ Gonihedric Hamiltonian, $H_{\kappa=0}$, following the suggestion of \cite{5}. The transition in this case is now first order
and hence no longer in the universality class of of the $2D$ Ising model, but the distinctive planar flip symmetry remains. 
We concluded that  the $M^{x,y,z}_{abs}$  of equ.\,(\ref{Mabsxyz}) might serve as an order parameter for the isotropic model and found that the  choice of reference direction $x,y,z$ was immaterial in this case. The absence of non-trivial structure in the distribution $P(M_{2D})$ of the coplanar magnetizations suggested that such a global parameter was sufficient to capture the order.

Somewhat unexpectedly, the standard magnetic susceptibility $\chi$  also shows the characteristics of an order parameter in the simulations of the isotropic Gonihedric model since it was measured to be
one at  high temperatures and dropped sharply to zero at the pseudo-critical point. However, $\chi$  failed to distinguish between an ensemble of flipped toy ground states and purely random configurations. We suggested that the behaviour in the full Gonihedric simulations was a dynamical consequence of freezing and that a multicanonical simulation might help to clarify this.

Regarding similar order parameters in other models, the dual Gonihedric model may be written as strongly anisotropic Ashkin-Teller model and also possesses similar flip symmetries to the purely plaquette model  \cite{99}. Understanding the low temperature order in this dual model, whose standard magnetization and polarization are zero at all $\beta$, might proceed along similar lines to those considered here.

\section{Acknowledgements}
D. A. Johnston would like to thank Martin Weigel, Wolfhard Janke and Adam Lipowski for helpful discussions.



\bigskip
\bigskip
\bigskip




\begin{thebibliography}{}

\bibitem{1} G. K. Savvidy and F.J. Wegner, Nucl. Phys. {\bf B413} (1994) 605.\\
            G. K. Savvidy and K.G. Savvidy, Phys. Lett. {\bf B324} (1994) 72.\\ 
            G. K. Savvidy and K.G. Savvidy, Phys. Lett.{\bf B337} (1994) 333.\\
            G.K.Bathas, E.Floratos, G.K.Savvidy and K.G.Savvidy, Mod. Phys. Lett. {\bf A10} (1995) 2695.\\
            G. K. Savvidy, K.G. Savvidy and F.J. Wegner, Nucl. Phys. {\bf B443}
            (1995) 565.\\
            G. K. Savvidy and K.G. Savvidy, Mod. Phys. Lett. {\bf A11} (1996) 1379.\\
            G. K. Savvidy, K.G. Savvidy and P.G. Savvidy, Phys.Lett. {\bf A221} (1996) 233.\\
     D. Johnston and R.K.P.C. Malmini, Phys. Lett. {\bf B378} (1996) 87.\\
     G. Koutsoumbas, G. K. Savvidy and K. G. Savvidy, Phys.Lett. {\bf B410} (1997) 241.\\
            J.Ambj\o rn, G.Koutsoumbas, G.K.Savvidy, Europhys.Lett. {\bf 46} (1999) 319.\\
            G.Koutsoumbas and G.K.Savvidy, Mod.Phys.Lett. {\bf A17} (2002) 751.
           

\bibitem{2} R.V. Ambartzumian, G.S. Sukiasian, G. K. Savvidy
            and K.G. Savvidy, Phys. Lett. {\bf B275} (1992) 99.\\
            G. K. Savvidy and K.G. Savvidy, Int. J. Mod. Phys.
            {\bf A8} (1993) 3393.\\
            G. K. Savvidy and K.G. Savvidy, Mod. Phys. Lett.
            {\bf A8} (1993) 2963.\\
            J. Ambj\o rn, G.K. Savvidy and K.G. Savvidy, Nucl.Phys. {\bf B486} (1997) 390.


\bibitem{3a} A. Lipowski J. Phys. {\bf A30} (1997) 7365.

\bibitem{3}   M. Baig, D. Espriu, D. Johnston and R.K.P.C. Malmini, J. Phys. {\bf A30} (1997)  405.\\
             A. Lipowski and D. Johnston, J. Phys. {\bf A33} (2000) 4451.\\
             A.Lipowski and D.Johnston, Phys.Rev. {\bf E61} (2000) 6375.\\
             A. Lipowski, D. Johnston and D. Espriu, Phys Rev. {\bf E62}  (2000) 3404.\\
             M. Swift, H. Bokil, R. Travasso and A. Bray, Phys. Rev. {\bf B62} (2000) 11494.\\
             S. Davatolhagh, D. Dariush and L. Separdar, Phys Rev. {\bf E81} (2010) 031501.


\bibitem{4} R. Pietig and F. Wegner, Nucl.Phys. {\bf B466} (1996) 513.\\
              R. Pietig and F. Wegner, Nucl.Phys. {\bf B525} (1998) 549.     


\bibitem{5} Y. Hashizume and M. Suzuki, Int. J. Mod. Phys. {\bf 25} (2011) 73.\\
Y. Hashizume and M. Suzuki, Int. J. Mod. Phys. {\bf B 25} (2011) 3529.


\bibitem{6} M. Suzuki, Phys.Rev.Lett. {\bf 28} (1972) 507.

\bibitem{7} T. Jonsson and G.K. Savvidy, Nucl.Phys. {\bf B575} (2000) 661;
Phys.Lett. {\bf B449} (1999) 253. 

\bibitem{8} C. Castelnovo, C. Chamon and D. Sherrington, Phys. Rev. {\bf B 81} (2010) 184303.

 
\bibitem{Marco} M. M\"uller, ``Multicanonical Analysis of the
Gonihedric Ising Model and its Dual'', Diploma thesis, University of Leipzig (2011) 

\bibitem{Fixed} M. Baig, J. Clua, D.A. Johnston and R. Villanova, Phys. Lett. 
{\bf B585} (2004) 180.

\bibitem{99}  D. Johnston and R.K.P.C.M. Ranasinghe, J. Phys. {\bf A44} (2011) 295004.




\end{thebibliography}
\end{document}